\documentclass[prb,aps,floatfix,showpacs,preprint,superscriptaddress]{revtex4}
\usepackage{amssymb}
\usepackage{epsfig}
\usepackage{graphicx}
\usepackage{dcolumn}
\usepackage{bm}
\usepackage{graphicx,amscd,amsmath,amssymb}
\usepackage[dvips]{hyperref}
\usepackage{textcomp}
\usepackage[OT1,T1]{fontenc}

\begin{document}

\title{Chiral Symmetry and Electron Spin Relaxation  of  
Lithium Donors in Silicon}

\author{V.~N.~Smelyanskiy}
\email{Vadim.N.Smelyanskiy@nasa.gov} \affiliation{NASA Ames Research Center, Mail Stop 269-3, Moffett
Field, CA 94035, USA}
\author{A.~G.~Petukhov}
\email{Andre.Petukhov@sdsmt.edu}
 \affiliation{Department of Physics, South Dakota School of Mines and Technology, Rapid City, SD
57701, USA}
\author{A.~M.~Tyryshkin}
\affiliation{Department of Electrical Engineering, Princeton University, Princeton, NJ 08544, USA}
\author{S.~A.~Lyon}
\affiliation{Department of Electrical Engineering, Princeton University, Princeton, NJ 08544, USA}
\author{T.~Schenkel}
\author{J.~W. Ager}
\affiliation{Lawrence Berkeley National Laboratory, Berkeley, CA 94720, USA}
\author{E.~E. Haller}
\affiliation{Lawrence Berkeley National Laboratory, Berkeley, CA 94720, USA}

\begin{abstract}
We report theoretical and experimental studies of the longitudinal
electron spin and orbital relaxation time of interstitial 
Li donors in $^{28}$Si. 
We predict 
that despite the near-degeneracy of the ground-state manifold 
the spin relaxation times 
are extremely long  for the
temperatures below 0.3 K. This prediction is based on a new finding of the
chiral symmetry of the donor states, which presists in the presence of random
strains and magnetic fields parallel to one of the cubic axes.
Experimentally observed kinetics of magnetization reversal at
2.1~K and 4.5~K are in a very close agreement with the theory.
To explain these kinetics we introduced 
a new mechanism of  spin decoherence based on  a combination of a small off-site
displacement of the Li atom and an umklapp phonon process. Both these factors weakly break
chiral symmetry and enable the long-term spin relaxation.
\end{abstract}
\pacs{03.67.Lx, 76.30.-v, 71.55.Cn, 71.55.Cn} 
\maketitle

\section{Intorduction}
Recent progress in building devices for quantum information processing (QIP) has lead to renewed interest
in studying the relaxation times of electron spins bound to impurities and in quantum dots to be used for
encoding quantum information \cite{Loss,kane1998a,Vrijen,Friessen,Petta}.   Electron spins bound to
shallow donors in Si are characterized by particularly long relaxation times, since the spin orbit
interaction is weak and isotopically enriched  $^{28}$Si is available~\cite{Gordon:1958,Tyryshkin:2003}.
However a weak interdonor spin coupling  represents a challenge for the controlled interaction of quantum
bits.

The interstitial lithium donor in silicon is unique amongst shallow donors because  it has an inverted
level structure and near degeneracy of the 1$s$ ground state manifold \cite{Aggarwal:65,Jagannath:1981}.
The latter gives rise to a strong long-range elastic-dipole coupling of the neutral Li donor {\it orbital}
states \cite{Smelyanskiy:2005}, similar to that obtained earlier for acceptors in Si \cite{Golding:2003}
but absent in the case of other donors.   In \cite{Smelyanskiy:2005} a quantum computing architecture was
proposed based on the stress-defined orbital qubits  and the above interaction was used to enable 2-qubit
gates. At the same time the near degeneracy of 1$s$ manifold in Si:Li also gives rise to a nontrivial
coupling between the spin and orbital degrees of freedom first investigated by Watkins and Ham
\cite{Watkins:70}. One can expect that the inverted level structure and near degeneracy in Si:Li will lead
to a very short electron spin relaxation times  which can also strongly affect orbital relaxation and be
detrimental for QIP. Somewhat surprisingly, in view of extensive studies of spin relaxation of other
shallow donors in silicon since the early sixties, both experimental
\cite{Gordon:1958,Feher:1959,Castner:63,Tyryshkin:2003} and theoretical
\cite{Hasegawa:60,Roth:60}, the question about Li spin relaxation has not been addressed.

\begin{figure}[tbh]
\includegraphics[width=0.74\linewidth,clip=true]{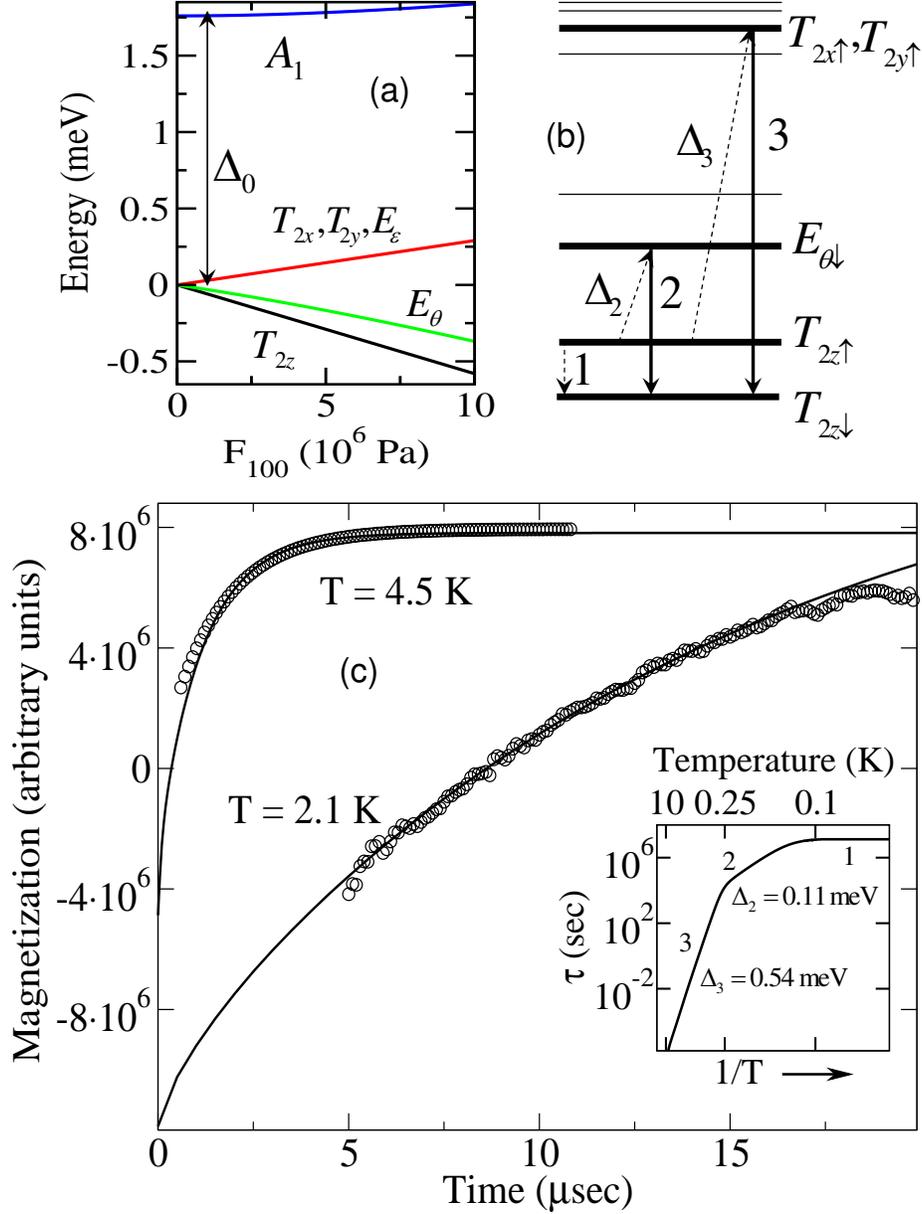}
\caption{ (a) Theoretical stress dependence of Si:Li 1$s$ energy levels at $B=0$. (b) Energy levels at finite
stress and $B\ne 0$. Solid lines 2,3
indicate the relatively fast one-phonon transitions connecting  the states with the {\it same} chirality.
Dashed line 1 indicates the slow spin-flip transition between the ``left'' and ``right'' states; 
(c) Comparison of
experimental and
 theoretical kinetics of magnetization recovery at
2.1 K and 4.5 K. The theoretical curves (thin solid lines) correspond to $V_\theta=V \sin\theta= 7\times
10^{-4}$~meV and
 $F_{100}=9\times 10^6$~Pa. The lower inset displays  the theoretically calculated
temperature dependence of the spin lifetime  $\tau=1/\lambda_{\rm min}$, an inverse of the lowest
eigenvalue of relaxation operator ${\cal L}$, Eq.~(\protect{\ref{relax}}). It is characterized by two activation energies $\Delta_{2}$
and $\Delta_{3}$ (the corresponding activation processes are shown in Fig.~\ref{fig:comparison}(b) with
dashed lines) \label{fig:comparison}}
\end{figure}

In this work we propose a new relaxation mechanism for coupled donor electron spin and orbital degrees
of freedom. It is based on (i) our new finding of the chiral symmetry of the donor states
that persists in prsence of the random starin and magnetic field of certain orientations  and 
(ii) an interplay between the umklapp phonon processes and a short-range electric
dipole interaction that only weakly break this symmetry and enable spin relaxation.  
As a result Li electron spin and
orbital relaxation times become very long in the presence of modest external stress ($\sim $ 10$^7$ Pa) at
temperatures below 0.3 K.  We also report pulsed ESR measurements of the longitudinal
spin relaxation time T$_1$ in $^{\rm 28}$Si:Li at temperatures 1.8 to 4.5~K. We find that in the absence
of external stress these  times are shorter than our instrumental resolution 
($<$ 100 ns) at temperatures
as low as 1.8 K. This rapid relaxation is caused by the degeneracy in the ground state manifold.  However,
a modest applied stress lifts the degeneracy, and leads to a much longer relaxation, as will be discussed
below. The results of the theoretical analysis of the selection rules and spin-orbital relaxation
mechanisms are in excellent  quantitative agrement with the experiment as shown in
Fig~\ref{fig:comparison}.

\noindent

The  ground state $1s$ manifold of an interstitial lithium donor in silicon can be described by the
Hamiltonian in a twelve dimensional Hilbert space formed by a tensor product of electron spin states
$|\chi_{\pm} \rangle$ and six 1$s$ orbitals corresponding to the six conduction  band minima ${\bf k}_i$
\cite{Watkins:70}
\begin{equation}H=H^{(0)}+H_{\rm d},\quad H^{(0)}=H_{\rm o}+H_{\rm B}+H_{\rm st}+H_{\rm so}.\label{Htot}
\end{equation}
\noindent The  term  $H_{\rm o}$  is the  spin-diagonal, effective mass orbital  1$s$ Hamiltonian also
describing the valley-orbit splitting due to the short range tetrahedral potential. The rest of individual
terms in $H$ to be considered below describe the   coupling of the donor electron
 to  magnetic field ($H_{\rm B}$), long-wavelength strain field ($H_{\rm st}$) , spin-orbit
interaction ($H_{\rm so}$), and a small off-center displacement of Li core from the symmetry site ($H_{\rm
d}$).

 For each of the two spin
orientations the six eigenstates of $H_{\rm o}$ are linear superpositions of the six silicon valley
orbitals, $\psi_{\mu}({\bf r})=\sum_{j=\pm 1}^{\pm 3}\alpha_{j}^{\mu}F_{j}({\bf r})u_{j}({\bf r})$. Here
$F_{j}({\bf r})$ and $u_{j}({\bf r})$ are, respectively, the hydrogenic envelope  and the Bloch function
for the j-th conduction band minima ${\bf k}_j=k_0 {\hat x}_j$ (${\hat x}_j$=$\pm {\hat x}, \pm {\hat y}$
or $\pm{\hat z}$). Eigenstate $\psi_{\mu}({\bf r})$ belongs to the irreducible representation $\mu$ of the
tetrahedral group characterized by the valley-orbit coefficients $\alpha_{j}^{\mu}$. In contrast to the
Group V substitutional donors the ground state of $H_{\rm o}$ is five-fold degenerate for each  spin
orientation. It is composed of a triplet of states $T_{2x}, T_{2y}, T_{2z}$  which are antisymmetric
combinations of just two opposite valleys, $\alpha_{j}^{\mu}=-\alpha^{\mu}_{-j}$, and a doublet
$E_{\theta}, \,E_{\epsilon}$ that are symmetric combinations of valleys,
$\alpha_{j}^{\mu}=\alpha^{\mu}_{-j}$, with a second-order zero at the origin. A fully symmetric singlet
state $A_1$ lies at $\Delta_0$=  1.76 meV above the $1s(E+T_2)$ orbital ground
state~\cite{Aggarwal:65,Jagannath:1981}, so that  matrix elements $\langle \psi_{\mu}|H_{\rm
o}|\psi_{\mu^{\prime}}\rangle=\Delta_0\delta_{\mu,A_1}\delta_{\mu^{\prime},A_1} \hat I$, where $\hat I$ is
2$\times$2 unit matrix operating in spin subspace.

\section{Parity-selection rules for acoustic-phonon transitions}
Electron-phonon interaction $H_{{\bf q}\nu}$ within 1$s(E$+$T_2)$ manifold can be  described in terms of
deformation potentials because the states there decay via emission of long-wavelength acoustic phonons
with $q$$\sim$10$^{5} \textmd{cm}^{-1}$. A matrix element between a pair of  donor states $\psi_\mu$ and
$\psi_\mu^{\prime}$  of  $H_{{\bf q}\nu}$  with phonon wavevector $q$  and polarization $\nu$ obeys parity
selection rules \cite{Castner:63} and can be written in the form
\begin{equation}
\langle \psi_\mu|H_{{\bf q}\nu}|\psi_{\mu^{\prime}}\rangle= \hat I\,\sum_{j=1}^{3}M^{j}_{{\bf q}\nu}
[A^{+}_{j,\mu,\mu^{\prime}}+q_j r_u
 A^{-}_{j,\mu,\mu^{\prime}}],   \label{umklapp}\end{equation}
 \noindent
 where $ A^{\pm}_{j,\mu,\mu^{\prime}}=
 \alpha_{j}^{\mu}\alpha_{\pm j}^{\mu^{\prime}}\pm\alpha_{-j}^{\mu}\alpha_{\mp j}^{\mu^{\prime}}$.
 Here  $M^{j}_{{\bf q}\nu}$ is the energy shift of
valleys ${\bf k}_{\pm j}$  due to the phonon mode ${\bf q}$ with polarization $\nu$. In can be expressed
in terms of the deformation potential constants  and angular functions \cite{Castner:63}. $A^{+}$ and
$A^{-}$ are zero for the states $\mu,\mu^{\prime}$ with the same and opposite parities, respectively. In
the latter case the transition amplitude (\ref{umklapp}) contains a very small quantity $q r_u \sim
10^{-4}$ where $r_u=\int F_z^2({\bf r})z\sin(\kappa_u z)d{\bf r}$ is determined by the overlap in
$k$-space  of the envelope functions localized near the different conduction band minima separated by
umklapp vector $\kappa_u \simeq 0.3 \times 10^{8}$ cm$^{-1}$ \cite{Castner:63} .

Matrix elements $\langle \psi_\mu|H_{\rm st}|\psi_{\mu'}\rangle=\Xi_u\hat I\sum_{j=1}^{3}A^{+}_{j\mu\mu'}
u_{jj}$ depend on a shear deformation potential $\Xi_u$ that  couples  the long-wavelength strain tensor
components $u_{jj}$ to the donor states \cite{Watkins:70}.  For
 most experimental situations the strain is caused by a superposition of an
external uniaxial and much weaker random biaxial stress that completely removes the orbital degeneracy. A
compressive uniaxial (001) stress $F_{001}$  only  partially splits the
 1$s(E+T_2)$ manifold \cite{Watkins:70,Jagannath:1981} as shown in
Fig.~\ref{fig:comparison}. The ground state is $T_{2z}$ and the first excited state is  a superposition of
$E_\theta$ and $A_1$ orbitals with dominant $E_\theta$ character. The third  level is triple-degenerate,
corresponding to states $T_{2x}, T_{2y}, E_{\epsilon}$ as shown in Fig.~\ref{fig:comparison}.  For
$F_{001} =9$$\times$10$^{6}$ Pa the splitting of $T_{2z}$ and $E_\theta$ levels is $\simeq$ 0.2 meV; much
larger than the random-stress induced level broadening in $^{28}$Si ~\cite{Thewalt:03,Yang:06}. We note
that due to the symmetry of an interaction involving deformation potentials \cite{Watkins:70} an
\textit{arbitrary}  biaxial stress  mixes only even-parity states, $A_1,E_\theta$ and $E_\epsilon$ and
therefore preserves the above parity-selection rules.

Matrix elements of the  Zeeman Hamiltonian equal
\begin{equation} \langle \psi_\mu|H_{\rm B}|\psi_{\mu'}\rangle=
\sum_{j=1}^{3}g_{\bot} \mu_B B_j \hat S_j(\delta_{\mu \mu'}+ \varepsilon A^{+}_{j\mu\mu'}),\label{B}
\end{equation}
\noindent where $\mu_B$ is the Bohr
magneton, $\hat S_j$ and $B_j$ are, respectively, 
the 3 cartesian components of the electron spin
operator and magnetic field. Also $\varepsilon=(g_{\parallel}-g_{\bot})/g_{\bot}$, and $g_{\parallel},g_{\bot}$ are the components of the g-tensor \cite{Watkins:70}.

There exists a small spin-orbit interaction in the ground state of Li donor  induced by the host silicon
lattice \cite{Watkins:70}. It is typically smaller then the strain  and  Zeeman terms. The spin-orbit
Hamiltonian
$H_{\rm so}$ 
reads~\cite{Watkins:70}:
\begin{eqnarray}
H_{\rm so}=-i \hat S_z\left(\lambda^{\prime} | T_{2z} \rangle \langle E_{\varepsilon}| +
\lambda
|T_{2x}\rangle\langle T_{2y}|\right)\nonumber\\
-i \hat S_x\left(\lambda^{\prime} |E_{+}\rangle\langle T_{2x}|+
\lambda
|T_{2y}\rangle\langle T_{2z}|\right)\nonumber\\
-i \hat S_y\left(\lambda^{\prime} |E_{-}\rangle\langle T_{2y}|+\lambda
|T_{2z}\rangle\langle T_{2x}|\right)+\,\,\,\textrm{h.c.},\label{SO}
\end{eqnarray}
\noindent where
$|E_{\pm}\rangle \equiv
 \frac{1}{2}\left(|E_{\varepsilon}\rangle
 \pm\sqrt{3}|E_{\theta}\rangle\right)$,
and small constants $\lambda\simeq 0.003$ meV and $\lambda^{\prime}=0.007$ meV determine the strength
 of the spin-orbit coupling within the triplet and
 between the triplet and singlet, respectively~\cite{Watkins:70}.

\section{Chiral symmetry}

Spin-orbit Hamiltonian $H_{\rm so}$ mixes the orbital states of opposite parities and, as one might
expect, it violates the above parity selection rules for the transition rates. However, we show below that
the time-reversal symmetry in the absence of a magnetic field leads to a factorization of the total
Hamiltonian and emergence of the new selection rules that persist for finite magnetic fields of certain
orientations. We 
assume  that the magnetic field ${\bf B}=B {\hat z}$.
Then the
eigenstates of the Hamiltonian $H^{(0)}$ form two non-overlapping subspaces
with an element  in each subspace being a linear superposition of the following 6 basis
vectors $|\chi_{\sigma}\rangle |\psi_{\mu}\rangle$:
 \begin{eqnarray}
|\Phi_{\sigma}^{(0)}\rangle&=&|\chi_{\sigma}\rangle\left(c_z |T_{2z}\rangle+c_\theta
 |E_\theta\rangle+c_\varepsilon |E_{\varepsilon}\rangle+c_{A_{1}}
 |A_{1}\rangle\right)\nonumber\\
&+&|\chi_{-\sigma}\rangle \left(c_x |T_{2x}\rangle+c_y |T_{2y}\rangle\right),\quad \sigma=\pm
1,\label{subspaces}
\end{eqnarray}
\noindent where  
$\hat \sigma_{z}|\chi_\sigma\rangle=\sigma |\chi_\sigma\rangle$, 
$\hat\sigma_z$ is the Pauli matrix, and $\sigma$ is
 a subspace index. The eignstates $|\Phi_{+}^{(0)}\rangle$ and their 
time-reversed counterparts $|\Phi_{-}^{(0)}\rangle$ are also the eigenstates
of the ``chirality'' operator $\hat Z = \hat R_z(\pi)\hat\sigma_z$, where
$\hat R_z(\pi)$ is the operator of rotation through the angle $\pi$ around
$z$-axis. Indeed, the operator $\hat Z$ commutes with the Hamiltonian 
$H^{(0)}$ and
$\hat Z|\Phi_{\pm}^{(0)}\rangle = \pm|\Phi_{\pm}^{(0)}\rangle$. 
In the absence of the magnetic field 
pairs of similar eigenstates with opposite chiralities 
form Kramers doublets with the same energy. A
finite magnetic field $B$  lifts the Kramers degeneracy.  
However for $H_B$ with ${\bf B}=B {\hat z}$ the chirality still
remains a good quantum number and the subspaces do not mix despite
an arbitrary biaxial strain present in $H^{(0)}$.
Similar subspaces can be constructed for ${\bf B}=B
{\hat x}$ and ${\bf B}=B { \hat y}$.

Watkins and Ham postulated   a small displacement ${\bf d}$ of  the Li atom from a tetrahedral
interstitial site \cite{Watkins:70}, preferentially in the direction of the  uniaxial stress ${\bf
F}\parallel {\bf d}$ to generate the correct number of lines in the Si:Li ESR spectrum. It  gives rise to
a short-range electric dipole potential $H_{\rm d}$ (\ref{Htot}) with the strength $V$  and matrix
elements
\begin{equation}
 \langle\psi_\mu|H_{\rm d}|\psi_{\mu^\prime}\rangle = 2i V\hat I \sum_{j=1}^3 A^{-}_{j,\mu,\mu^\prime}
 d_j/d,\quad d_j={\bf d}\cdot \hat x_j,\label{dip}
\end{equation}
\noindent connecting  states  of opposite parities and the same spin orientation.  Using
(\ref{subspaces}), (\ref{dip}) and coefficients $\alpha^{\mu}_{j}$ one gets
\begin{equation}
\langle\Phi_{\sigma}^{(0)}|H|\Phi_{-\sigma}^{(0)}\rangle=0\quad {\rm for}\quad {\bf B}\parallel {\bf
d}\parallel \hat x_j,\label{PHP}
\end{equation}
where $\hat x_j={\hat z}$, ${ \hat y}$ or ${\hat x}$.   However, we will assume a small random
misalignment, $|\hat x_j\times{\bf d}/d|=
 |\sin\theta|\ll 1$ (perhaps, due to a random biaxial stress in addition to a strong uniaxial stress ${\bf F}$=$F \hat x_j$).
 It will lead to a slight mixing of the ``left'' and ``right'' states
$\langle\Phi_{\sigma}^{(0)}|H|\Phi_{-\sigma}^{(0)}\rangle\propto V_\theta\equiv V \sin\theta$.

\section{Transition rates}

Consider ${\bf B}$$\parallel$${\bf F}$$\parallel$$\hat x_j$ ($\hat x_j={\hat z}, { \hat y}\, {\rm or}\,
{\hat x}$). Then the
 matrix elements of  $H_{q\nu}$  between a pair of exact eigenstates, $H\Phi_n={\cal E}_n\Phi_n$, that are predominantly ``left'' and ``right'' have the form:  $ \langle
\Phi_{n}|H_{{\bf q}\nu}|\Phi_{m}\rangle \simeq C_{\bf q\nu}^{\rm u} q r_u+ C_{\bf q\nu}^{\rm
d}\frac{V_\theta}{{\cal E}_n-{\cal E}_m}$. The first term  is due to an umklapp process ( $q
r_u$$\sim$10$^{-4}$) and the second term is due to a small  misalignment $V_\theta$=$V\sin \theta$ of
${\bf d}$ and $\hat x_j$. Both terms are small and \textsf{ acoustic phonon transitions between the states of opposite chiralities are suppressed}. As shown below, under the appropriate  conditions  each  of the chiral
 subspaces can  be made nearly \lq\lq decoherence free".

 An important application of the above rule is a suppression of the spin-flip
transitions between the
 states  with  dominant characters $|\chi_{\pm 1}\rangle\otimes|\psi_\mu\rangle$
 that differ only in spin orientation. Those states always have opposite chiralities
\begin{figure}[tbh]
\includegraphics[width=0.8\linewidth,clip=true]{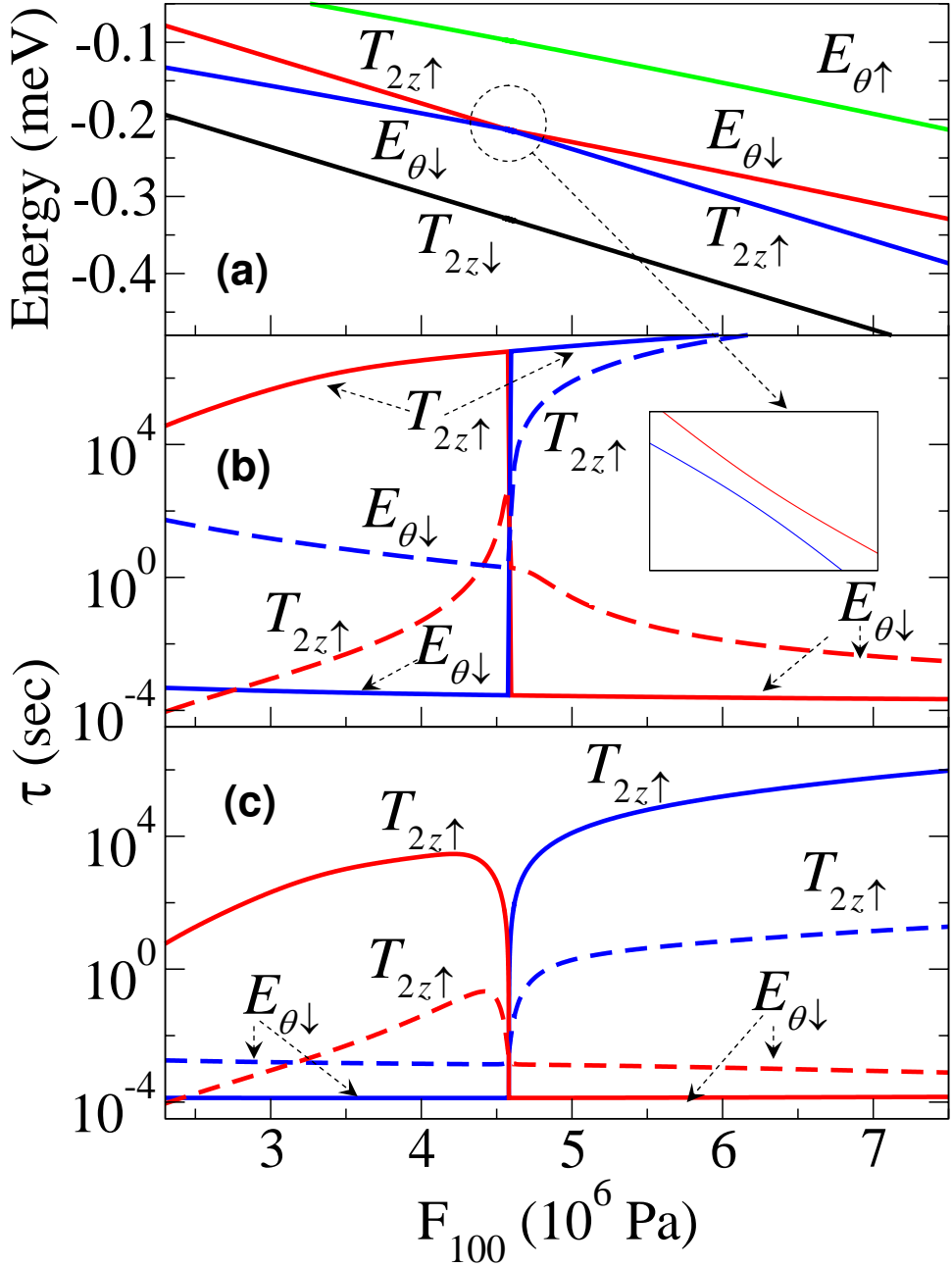}
\caption{(Color online.) (a) Calculation of the first four energy levels of Si:Li donor in a magnetic
field $B=1$~T; (b) calculation of the lifetimes of the first two excited states  versus compressive 001
stress at $T$=0.1~K, $B=1$~T, and $V_\theta=V\sin\theta=0$. 
Solid lines --- ${\bf B}\parallel \hat z$ and dashed lines --- ${\bf B}\parallel \hat
x$. Inset shows magnified region of the avoided crossing for the magnetic field ${\bf B}\perp \hat z$;
(c) calculation of the lifetimes of the first two excited states  versus compressive 001
stress at $T$=0.1~K, $B=1$~T, and $V_\theta=V\sin\theta= 4.8\times 10^{-4}$~meV.
Solid lines --- ${\bf B}\parallel \hat z$ and dashed lines --- ${\bf B}\parallel \hat
x$.} \label{fig:energy}
\end{figure}
We shall label each eigenstates after its dominant character and consider ${\bf B}=B\hat z$, ${\bf
F}=F_{001}\hat z$. If the compressive stress $F_{001}$ is sufficiently large the Zeeman splitting $g_\bot
\mu_B B$ becomes smaller than a stress-induced splitting between $E_{\theta}$ and $E_{T_{2z}}$ (see
Fig.~\ref{fig:energy}). Then at
low temperature the state $|T_{2z}\hspace{-0.04in}\uparrow\rangle$ decays directly into the state
$|T_{2z}\hspace{-0.04in}\downarrow\rangle$ of the opposite chirality. There is a  competition between
umklapp and dipole-induced  processes. For moderate uniaxial stress and magnetic field the dipole
contribution to the lifetime $\tau_{\rm T_{2z}\hspace{-0.02in}\uparrow}$ dominates
\begin{equation}
\tau^{-1}_{\rm T_{2z}\hspace{-0.02in}\uparrow}\simeq  \tau_{A_1}^{-1} \left(\frac{ \lambda
\lambda^{\prime} V_d\sin\theta_d}{\Delta_{0}^{3}}\right)^2
\frac{41\upsilon^3}{x^8}\left(\left(\frac{x}{3}\right)^3 +\upsilon^2\right).\label{Tzup1}
\end{equation}
\noindent Here $\upsilon=g_{\bot}\mu_B B/\Delta_0$  and $x=\Xi_u(S_{12}-S_{11})F_{001}/6\Delta_0$ where
 $S_{11},S_{12}$ are elastic compliance moduli. Prefactor in (\ref{Tzup1}) corresponds to the lifetime $\tau_{A_1}\sim$1
psec of the states  $A_{1}$  that decay within the same subspaces to    $E_{\theta}$ and $E_{\epsilon}$.

For greater stress and/or magnetic field the umklapp process dominates the decay, giving
\begin{equation}
\tau^{-1}_{\rm T_{2z\uparrow}}=\tau_{A_1}^{-1}\,\xi_u\,\left(\frac{\lambda\lambda^{\prime}\Delta_0
a_{\parallel}}{\hbar u_{\perp}}\right)^2 \, \frac{7\upsilon^5}{3(x^2-\upsilon^2)^2}. \nonumber
\end{equation}
\noindent Here $a_{\parallel}$ is the
 longitudinal Bohr radius, $u_\bot$ is the transverse sound
velocity, and $\xi_u\simeq 5\cdot 10^{-5}$ depends on the overlap factor $r_u$ given above via
$(r_u/2a_{\parallel})^2=\xi_u-\xi_{u}^{5/6}$.

We calculated transition rates $W_{n\rightarrow m}$ between each pair of the 12 eigenstates
$|\Phi_n\rangle$ by numerical diagonalization of $H$ (\ref{Htot}) and  employing
 the  electron-phonon interaction (\ref{umklapp}). The stress
dependence of the lifetimes of the first two excited states of Si:Li donors at low temperatures is shown
in Fig.~\ref{fig:energy} for ${\bf B}\parallel \hat z$ and ${\bf B}\parallel \hat x$. These states
can be classified as either orbital (predominantly $E_{\theta\downarrow}$) or spin (predominantly $T_{2z\uparrow}$)
excitations. If the stress-induced splitting (SIS) is smaller than the Zeeman splitting (ZS) 
the first and second excited states correspond to the orbital and spin excitations respectively. 
As soon as SIS overcomes ZS the two states switch their characters. 

The orbital and spin excitations have distinctly different stress dependencies 
of their lifetimes.
The lifetime of the orbital exitation is decreasing with the stress increase while the spin excitation 
behaves quite oppositely. The stress increase leads to a large separation between 
$E_{\theta\downarrow}$ and $T_{2z\downarrow}$ states. Consequently the probability of 
$E_{\theta\downarrow} \rightarrow T_{2z\downarrow}$ phonon decay
grows as the phonon density of states or faster. On the contrary, the stress does not affect the 
Zeeman doublet
$T_{2z\uparrow}$---$T_{2z\downarrow}$ but rather separates it from other states of the manifold 
making the spin excitation $T_{2z\uparrow}$ more stable.   
The lifetime decrease of the orbital excitation is more pronounced for an ideal position of 
the Li-donor, i.e. for $V_{\theta}=0$ (see Fig.~\ref{fig:energy}(b)) and for ${\bf B}\parallel \hat x$.
In this case the orbital excitation and the ground state have different chiralities and the former may
decay through the umklapp process only. This adds additional stress dependence due to the 
matrix element of the umklapp process. The magnitude and the 
stress dependence of $E_{\theta\downarrow}$-lifetime shown in Fig.~\ref{fig:energy}(b) 
(dashed line) is very similar to our previous result~\cite{Smelyanskiy:2005}. The dipole
term $V_\theta$ reduces both the magnitude and the stress dependence of $E_{\theta\downarrow}$-lifetime, 
which is weakly decreasing with stress near the value of 10$^{-4}$~s (see Fig.~\ref{fig:energy}(c)).

\section{Experimental results and comparison with theory} 
We conducted a pulsed ESR study of the Li donor spin relaxation under
stress. An isotopically-enriched $^{28}$Si crystal (800 ppm residual $^{29}$Si) was $^7$Li doped by
implanting with a variable energy, up to 360~keV, and a total dose 6$\cdot$10$^{12}$ of $^7$Li/cm$^2$
followed by a 30 min anneal at 800~$^0$C to ensure a homogeneous distribution of the lithium. The
estimated Li concentration is $\sim$~10$^{14}$~cm$^{-3}$. Uniform, (110), in-palne tensile stress was
applied to the Si by gluing fused silica slides (1 mm thick) to the two (001) faces with Apiezon N vacuum
grease. This ``sandwich'' was then frozen to liquid helium temperatures.  The thermal expansion
coefficient of fused silica is substantially smaller than that of silicon in the temperature range 4-300
K~\cite{Nyilas2004}, and therefore freezing the sandwich results in an uniform in-plane tensile strain,
equivalent to the action of a compressive uniaxial
stress $F_{001}$ along the (001) axis of the silicon crystal (perpendicular to the faces). 
The spin resonance experiments were done at X-band ($\sim$0.34~T). Under $F_{001}$  the ESR spectrum is a
single line as described by Watkins and Ham~\cite{Watkins:70} though our linewidth (8~$\mu$T) is
considerably narrower than in their natural Si, since inhomogeneous broadening from the $^{29}$Si is
eliminated in the isotopically enriched $^{28}$Si \cite{footnote1}.  The  electron magnetization decay
measured at 2.1 and 4.5 K with a conventional inversion recovery pulse sequence~\cite{Schweiger2001} is
shown in Fig.~\ref{fig:comparison}.

The stress  $F_{001}$  in the sandwiched samples was estimated by performing Fourier transform infrared
spectroscopy (FTIR) with a Bruker IFS 66v/S spectrometer (resolution, 0.14 cm$^{-1}$) on a similarly
prepared stressed samples of Li-doped natural float zone silicon (a hole with a diameter 3 mm in the
silica slides allowed FIR transmission through the stressed sample). The  stress-induced splitting 0.202
meV was observed between the $1s(E+T_2)\rightarrow 2p_0(-)$ line at 21.287 meV and $1s(E+T_2)\rightarrow
2p_0$ line at 21.489 meV. Using this splitting we estimate $F_{001}\simeq 9.0\times 10^6$ Pa  based on the
previous measurements with calibrated stresses \cite{Jagannath:1981} and expressions for the
stress-induced line shifts \cite{Aggarwal:65,Jagannath:1981}.

To explain the magnetization recovery  data  in Fig.~\ref{fig:comparison}(b) we studied the multilevel
kinetics,
\begin{equation}  
\dot{\rho}_{ n}=\sum_{m=1}^{12}{\cal L}_{n m}\rho_{ m}, 
\end{equation}
where
\begin{equation}  
{\cal L}_{nm}= W_{m\rightarrow
n}-\delta_{n m}\sum_{k=1}^{12}W_{ n \rightarrow k}
\label{relax}
\end{equation} 
is a relaxation operator and $\rho_{n}=\langle
\Phi_n|\hat \rho|\Phi_n\rangle$ are individual state populations.  We used the stress value
$F_{100}=9\times 10^6$~Pa estimated from the above FTIR experiment. The  only unknown parameter in the
expressions for the decay rates $W_{nm}$  is $V_\theta$, or rather $V_{\theta}^{2}$ .  We were able to fit
the two experimental time traces at $2.1$~K and $4.5$~K using $V_\theta=7\times 10^{-4}$ meV ($\sim
0.25\lambda$) as shown in Fig.~\ref{fig:comparison}. This value corresponds to a statistical average
$(\langle V_{\theta}^{2}\rangle)^{1/2}$  over all  Li donors. We note the multi-exponential character of
the decay at 4.5 K. The temperature dependence of the long-time magnetization kinetics is determined by
that of the smallest eigenvalue of the relaxation operator $\lambda_{\rm min}(T)$.  The magnetization
decay proceeds via thermal activation to rapidly decaying excited states as shown in
Fig.~\ref{fig:comparison}(b). For $T$~$<$~0.1~K  the spin life-time
$\tau$ saturates at  large value consistent with those shown in Fig.~\ref{fig:energy}.

\section{Conclusions}
To conclude, we introduced a new spin-orbital  relaxation mechanism for Li donors in Si and predicted
theoretically that the longitudinal spin relaxation $T_1$ time of a Li donor in Si under the uniaxial
compressive stress and magnetic field aligned along the $\hat x$,$\hat y$ or $\hat z$ axis can be very
large (hours) for $T$~$<$~0.3~K despite the inverted level structure of interstitial Li donor as compared to
Group V substitutional donors in silicon. The orbital relaxation time is 0.1 msec at that temperature. By
virtue of the chirality selection rules for the electron-phonon interaction, the
donor electron spin   can decay only via  inter-valley umklapp
processes or a very weak direct phonon process caused by a small Li off-site displacement misalignment. At
higher temperatures $T_1$ is reduced to a $\mu$s range due to thermal population of the lower excited
states which in turn promotes fast transitions within the same chiral subspace allowed by the
parity-selection rules. We obtained a very close  (within few per cent) agreement with pulsed ESR
measurements of $T_1$ times at $T=2.1$~K (12~$\mu$s) and at $T=4.5$~K (1.5~$\mu$s) that corroborates our
prediction of extremely long-lived spin states of Li donors in Si at lower temperatures. Since the chiralty
is a universal property of the silicon host ur study can be
applied to substitutional donors in Si with similar near-degeneracy of the 1$s(E+T_2)$ coupled
to the ground state via 
stray electric fields and umklapp processes. Our results of the very long
Li spin lifetime can also stimulate the analysis of  the possibility of Li spin-based QIP applications with direct
long-range elastic-dipole interaction between the Li donor spins similar to that considered in
\cite{Golding:2003,Smelyanskiy:2005}.

\section{Acknowledgments}

This research was supported  by  US NSA,
NASA Grant NNX07AL35A, ONR Grant N00014-06-1-0616,  and DOE 
Contract No. DE-AC02-05CH11231.



\begin{thebibliography}{99}
\bibitem{footnote1} Li line is $\sim$100 times stronger than  Li$_2$O  line.
\end{thebibliography}
\end{document}